\documentclass[english,prd,twocolumn,superscriptaddress,nofootinbib,preprintnumbers]{revtex4}
\usepackage[latin1]{inputenc}
\usepackage{graphicx}
\usepackage{graphicx,subfigure}
\usepackage{amssymb}
\usepackage{epstopdf}

\usepackage{latexsym} 
\usepackage{feynmf}		

\makeatletter

\usepackage{amsfonts}
\usepackage{dcolumn}
\usepackage{hyperref}

\def\be{\begin{equation}}
\def\ee{\end{equation}}
\def\ba{\begin{eqnarray}}
\def\ea{\end{eqnarray}}
\def\bs{\begin{subequations}}
\def\es{\end{subequations}}

\usepackage{color}

\makeatother

\usepackage{babel}
\makeatother
\begin{document}

\title{Sub-eV scalar dark matter through the super-renormalizable Higgs portal}

\author{Federico Piazza}
\affiliation{Perimeter Institute for Theoretical Physics, Waterloo, ON, N2L 2Y5, Canada }
\affiliation{Canadian Institute for Theoretical Astrophysics (CITA), Toronto, Canada }
\email{fpiazza@perimeterinstitute.ca}

\author{Maxim Pospelov}
\affiliation{Perimeter Institute for Theoretical Physics, Waterloo, ON, N2L 2Y5, Canada }
\affiliation{Department of Physics and Astronomy, University of Victoria, Victoria, BC,  V8P 1A1, Canada}
\email{mpospelov@perimeterinstitute.ca}

\begin{abstract}

The Higgs portal of the Standard Model provides the opportunity 
for coupling to a very light scalar field $\phi$ via the super-renormalizable 
operator $\phi(H^\dagger H)$. This allows for the existence of a very light scalar dark matter
 that has coherent interaction with the Standard Model particles 
and yet has its mass protected against radiative corrections. 
We analyze ensuing  constraints from the fifth-force measurements, along with the cosmological 
requirements. We find that the detectable level of the fifth-force can be 
achieved in models with low inflationary scales, and certain amount of fine-tuning in the 
initial deviation of $\phi$ from its minimum.

\end{abstract}

\date{\today}

\maketitle

\section{Introduction}

About 95\% of the energy budget of the Universe consists of 
"dark" -- and unknown -- components. This is a strong motivation for
considering and studying hidden sectors beyond the Standard Model (SM).
Gravitational effects of dark matter cannot reveal the mass of its 
constitutents, and indeed a wide variety of mass ranges, from the inverse galactic 
size to the super-Planckian scales, is conceivable. While many models that possess 
stable particles with masses comparable to the SM energy scales have been a 
subject of incessant theoretical and experimental activity, models with light 
sub-eV mass scale dark matter received far less attention. 

Below the eV mass scale the dark matter would have to be of integer spin, and be 
produced non-thermally. The only chance of detecting such dark matter non-gravitationally 
would occur if such particles are converted into electromagnetic radiation in the external fields or 
they modify the interaction stength of SM particles. 
But if light dark matter interacts with the SM, then immediately its lightness comes to question
as the quantum loops with SM particle may easily destabilize the mass scale. 
A prominent particle in this category is the QCD axion \cite{QCD} that interacts with the 
SM currents derivatively, $j_\mu\partial_\mu a $, and has its tiny mass
generated by the non-perturbative QCD effects protected at any loop level. 
Because of the pseudoscalar nature of $a$ and its derivative couplings,
it does not generate a long-range attractive force.

A very natural question to ask is whether SM allows for couplings to other types of sub-eV 
dark matter fields that lead to additional observable effects. 
For a recent review of the light sector phenomenology see, {\em e.g.} \cite{Ringwald}.
Real scalar field $\phi$ and the 
vector field $V_\mu$ provide such opportunities with their couplings to the 
SM fields via the so-called  Higgs and vector portals: 
\begin{eqnarray}
\label{portals}
(A \phi + \lambda \phi^2) H^\dagger H & & \qquad {\rm Higgs~portal}
\\
\nonumber
J_\mu V_\mu;~~ \partial_\mu J_\mu = 0 & & \qquad {\rm Vector~portal},
\end{eqnarray}
where $H$ is the Higgs doublet, $A$ and $\lambda$ are parameters and 
$J_\mu$ is some locally conserved SM current, such as hypercharge of baryon current. 
If there is some initial value for $\phi$ or $V_\mu$ fields with 
respect to their zero energy configurations, one can source part/all of the 
Universe's energy density from the coherent oscillations around the minimum. 

The perils of low mass scale stabilization are immediately apparent in Eq. (\ref{portals}). 
Indeed, any loops of the SM fields would tend to induce the correction to the 
mass of $\phi$ field $\sim \lambda \Lambda_{UV}^2$, where $ \Lambda_{UV}$
is the highest energy scale in the problem serving as the ultra-violet cutoff. 
Therefore, $\lambda$ should be taken to incredibly small values, making this 
portal irrelevant for the phenomenology of sub-eV dark matter. In contrast, the 
vector portals and the 
super-renormalizable Higgs portal, $A\phi H^\dagger H$, allow to avoid problems 
with technical naturallness. In the latter case loop corrections scale only as $A^2 \log\Lambda_{UV}$,
while the quadratic divergences affect only the term linear in $\phi$, which can typically be absorbed 
in an overall field shift. 
In this paper we examine generic consequences of this coupling
for the sub-eV scalar dark matter, leaving vector dark matter to future studies. 

\section{Super-renormalizable portal to the scalar dark matter}

The specific case of a singlet scalar $\phi$ coupled via a super-renormalizable 
term of the  type $\phi H^\dagger H$, (see {\em e.g.}~\cite{wil,wise,bar,ber,batell,ahlers} and references therein), 
has been mostly studied in connection with electroweak and GeV-scale phenomenology,
with a notable exception of \cite{ber,Pospelov}, where a possibility of super-weakly interacting 
Higgs-coupled dark matter was pointed out.  
The scalar potential in the model of interest reads as:
\begin{equation} \label{potential}
V = - \frac{m_h^2}{2}H^\dagger H + \lambda (H^\dagger H)^2 + A H^\dagger H\phi + \frac{m_\varphi^2}{2} \phi^2\, .
\end{equation}
This model is explicitly renormalizable and does not require any additional UV completion 
(if one is willing to tolerate the usual fine-tuning problem with $m_h^2$ itself). We chose to redifine away 
possible linear terms in $\phi$ by shifting the field, and absorbing $A\Delta \phi$ into $m_h^2$. 

After spontaneous symmetry breaking, the two fields acquire a vacuum expectation value, $\langle H^\dagger H \rangle = v^2/2$, $\langle \phi\rangle = \phi_0$, where
\begin{equation} \label{shifts}
v^2 = \frac{m_h^2}{2 \lambda - A^2/m_\varphi^2},\quad
\phi_0 = - \frac{A v^2}{2 m_\varphi^2}
\end{equation}
and $v = 246$ GeV. The potential (\ref{potential}) has a stable minimum only if $A^2/m_\varphi^2 < 2 \lambda$, which is what we assume in the following; otherwise, it develops a runaway direction in the $(\phi, H^\dagger H)$ plane
unless additional nonlinear $\phi^4$ terms
are introduced. The low energy dynamics is encoded in the two physical fields $h$ and $\varphi$, defined as 
\begin{equation}
H = \frac{1}{\sqrt{2}} \left(\begin{array}{c}
0\\
v+h\\
\end{array} \right), \qquad \phi = \phi_0 + \varphi
\end{equation}
and with Lagrangian
\begin{eqnarray}
{\cal L} &=& \frac{(\partial h)^2}{2} + \frac{(\partial \varphi)^2}{2} 
- \frac{m_h^2}{2} h^2 - \frac{m_\varphi^2}{2} \varphi^2 \\ & & - (A v) h \varphi  - \frac{A}{2} h^2 \varphi + \dots
\end{eqnarray}

As already noted, Higgs loops give only logarithmically divergent corrections to $m_\varphi$. Therefore, the 
requirement of technical naturalness bounds the 
scale of $m_\varphi$ from below by the coupling $A$. In summary, by defining the dimensionless ratio $x \equiv A/m_\varphi$, we assume $x\lesssim 1$ and $x < \sqrt{2 \lambda}$, although also values $x \ll 1$ will be considered. 

\section{Fifth force and Equivalence Principle violation}

\begin{figure}[b] 
\centering
\vspace{0cm}\rotatebox{0}{\hspace{-.5cm}\resizebox{.35\textwidth}{!}{\includegraphics{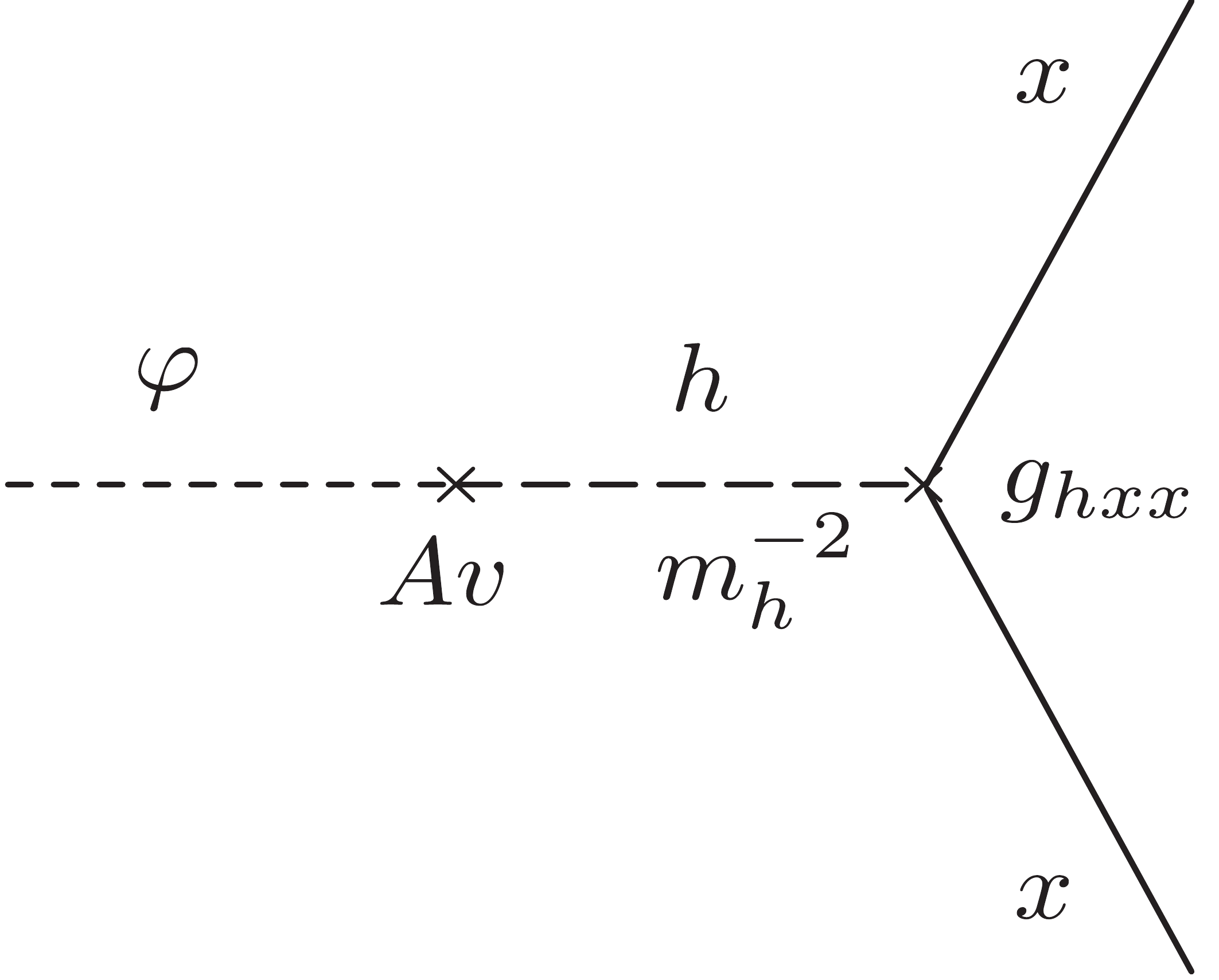}}}
\caption{The mixing with the Higgs $Av$ mediates the coupling of $\varphi$ to SM particles.\label{fig2}}
\end{figure}

The singlet $\varphi$ couples to SM particles through the mixing with the 
Higgs field. Depending on the  mass $m_\varphi$ and coupling $A$, 
the $\varphi$-mediated attractive force can produce testable deviations from 
$1/r^2$-gravitational force as well as composition dependence, thus violating the Equivalence Principle (EP).
The leading contributions to $\varphi$-couplings mediated by the $\varphi$-Higgs propagator is 
shown in Fig. \ref{fig2}. As a rule of thumb, the $\varphi$-couplings are suppressed 
with respect to the Higgs couplings by a factor of $A v/m_h^2$:
\begin{equation}
g_{\varphi x x} = \frac{A v}{m_h^2} g_{h x x} ,
\end{equation}
where $g_{h x x}$ is the effective dimensionless coupling of the Higgs to $x$-particle 
at very low momentum tranfer. Therefore, the effective Lagrangian describing the 
interactions with the SM gauge and fermion fields takes the following form:
\begin{equation} \label{couplings}
{\cal L}_{\rm eff} =  \frac{A v}{m_h^2}
\left(g_{h f f} {\bar f} f  + \frac{{ g}_{h \gamma \gamma}}{v} F_{\mu \nu} F^{\mu \nu} + \dots \right) \varphi \, .
\end{equation}

In the above, $g_{h f f}$ are the Yukawa couplings to fermions. Those can either be fundamental, as the SM couplings to quarks and leptons, $g_{h q q} = m_q/v$, $g_{h l l} = m_l/v$ where $m_q$ ($m_l$) is the mass of the quark (lepton) under consideration, or effective, as in the case of the nucleons. The latter includes the contributions from all heavy 
quarks contributing to the coupling to gluons 
$g_{hgg}$ that provide a dominant contribution in the chiral limit \cite{SVZ}.
Below the QCD scale, the estimate of  the effective Yukawa coupling of the Higgs 
to nucleons is rather uncertain due to a poorly known strangeness content of the nucleon 
in the $0^+$ channel:
\begin{equation} 
\label{nucleons}
g_{h N N} \simeq \frac{200-500 {\rm\,    MeV}}{v} \sim O(10^{-3}).
\end{equation}
This is much larger than the naive contribution of up and down quarks.

The violation of EP is evident from the fact that the electrons and nucleons have  
couplings to the $\varphi$ field that do not scale exactly with masses, 
\begin{equation}
  \frac{g_{hee}}{m_e} \neq \frac{g_{hNN}}{m_{\rm nuc}} .
\end{equation}

The effective coupling of the Higgs to the electromagnetic field, 
$g_{h \gamma \gamma}$, is obtained by integrating out 
heavy charged particles, and the question of which one is ``heavy" depends 
on the characteristic $q^2$ of (virtual) photons. 
The coupling  $g_{h \gamma \gamma}$ can be written in the following form 
(see, {\em e.g.} \cite{Higgs}):
\begin{equation}
g_{h \gamma \gamma} = \frac{\alpha_{\rm EM}}{6\pi}\left( 3 \sum_q Q_q^2 + \sum_l Q_l^2 -\frac{21}{4}\right),
\end{equation}
where summation goes over the quark and lepton fields with charges $Q_q$ and $Q_l$, and the last term is due to the 
the $W$-bosons. 
For the purpose of calculating the $\varphi\to \gamma \gamma$ decay, one has to sum over 
$e,\mu,\tau$ and $c,b,t$. Corrections coming from the light quark sector are subdominant,
because in the chiral limit they contribute at two-loops. In practice, their 
contribution would amount to at most 10\% correction. Including these fermion contributions 
gives $g_{h \gamma \gamma}(q^2=m_\varphi^2) \simeq \alpha_{\rm EM} /(8 \pi)$. For the purpose of calculating the 
coupling of $\varphi$ to nuclei when the EM fraction of energy is taken into account, 
electrons should not be included in the sum, and muon contribution should include a 
form-factor. We are not going to pursue this calculation, because it turns out that 
$g_{h \gamma \gamma} $ provides  a subleading contribution to the EP violation.

Field $\varphi$ mediates a fifth force of range $\sim m_\varphi^{-1}$. More precisely, at the Newtonian level of approximation, the total effective gravitational potential between two bodies $A$ and $B$ at relative distance $r$, presents a Yukawa contribution due to the interaction of the long range field $\varphi$,
\begin{equation}
V(r) = - G \frac{m_A m_B}{r}(1+ \alpha_A \alpha_B \, e^{- m_\varphi r})\, .
\end{equation}
 The scalar couplings $\alpha$ can be expressed in terms of the log-derivative of the masses as
\begin{equation}\label{alphaA}
\frac{\alpha_A}{\sqrt{2} M_P} = \frac{d \ln m_A(\varphi)}{d \varphi} ,
\end{equation}
where $M_P$ is the reduced Planck mass and $m_A(\varphi)$ includes terms in the Lagrangian that are bilinear in the fields and couple to $\varphi$, such as those in eq. (\ref{couplings}). When calculating $\alpha_A$, one should consider the leading universal contribution from the nucleons and all the corrections that are specific to the element $A$ (See 
{\em e.g.} ~\cite{dampol}). The main, species-independent part of the nuclear mass is given by 
$m_{\rm nuc}(N_A+Z_A)$, and the universal coupling $\alpha$ is 
obtained from eqs. (\ref{couplings}),  (\ref{nucleons}) and
(\ref{alphaA}): 
\begin{eqnarray} \label{alpha}
\alpha &=& g_{hNN} \frac{\sqrt{2} M_{P}}{m_{\rm nuc}} \frac{A v}{m_h^2}  \\ \nonumber &\simeq& 
10^{-3} \left(\frac{m_h}{115 \, {\rm GeV}}\right)^{-2} \frac{A}{10 ^{-8} \rm eV} .
\end{eqnarray}

\begin{figure}[htbp] 
  \begin{center}
     \subfigure{\label{fig2-a}\includegraphics[width=3.1in]{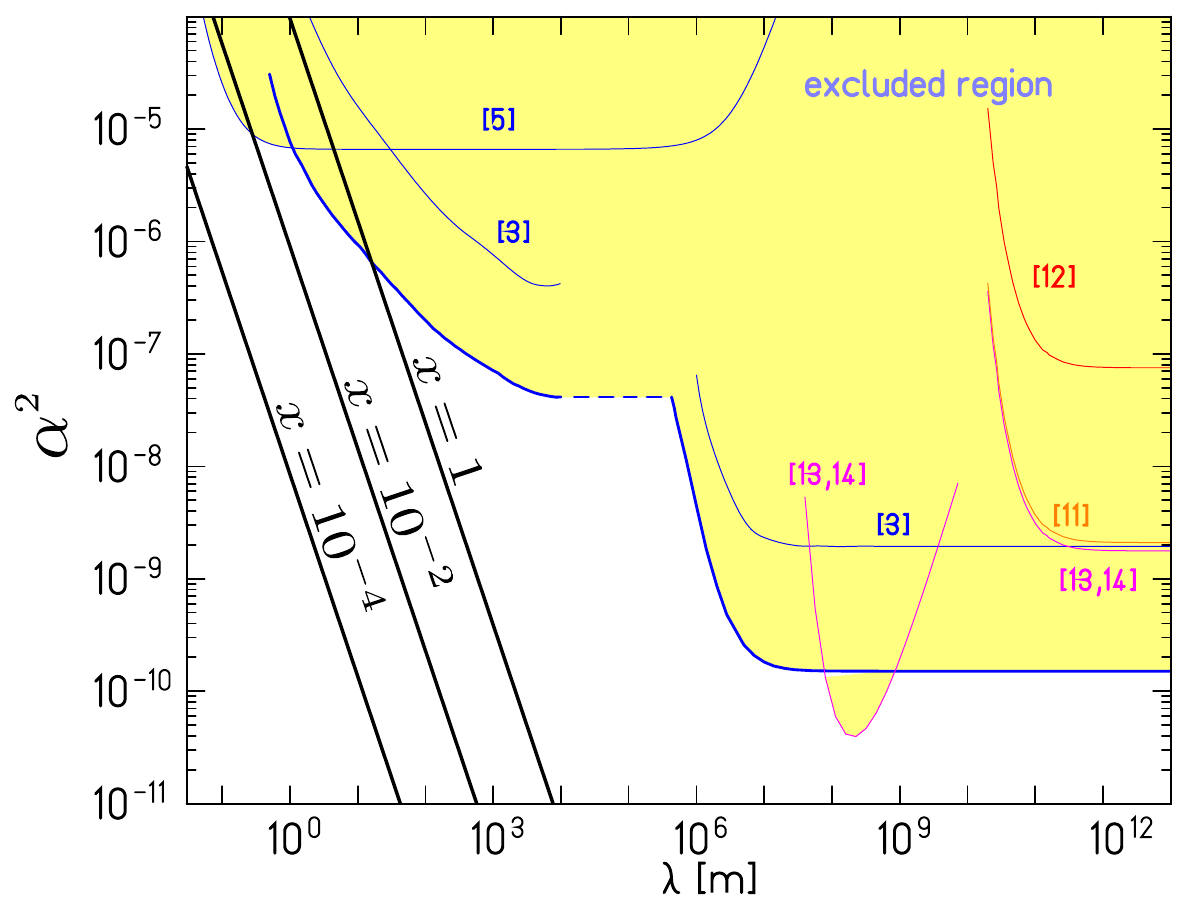}} \\ \vspace{-5mm} \hspace{2mm}
    \subfigure{\label{fig2-b}\includegraphics[width=3.23in]{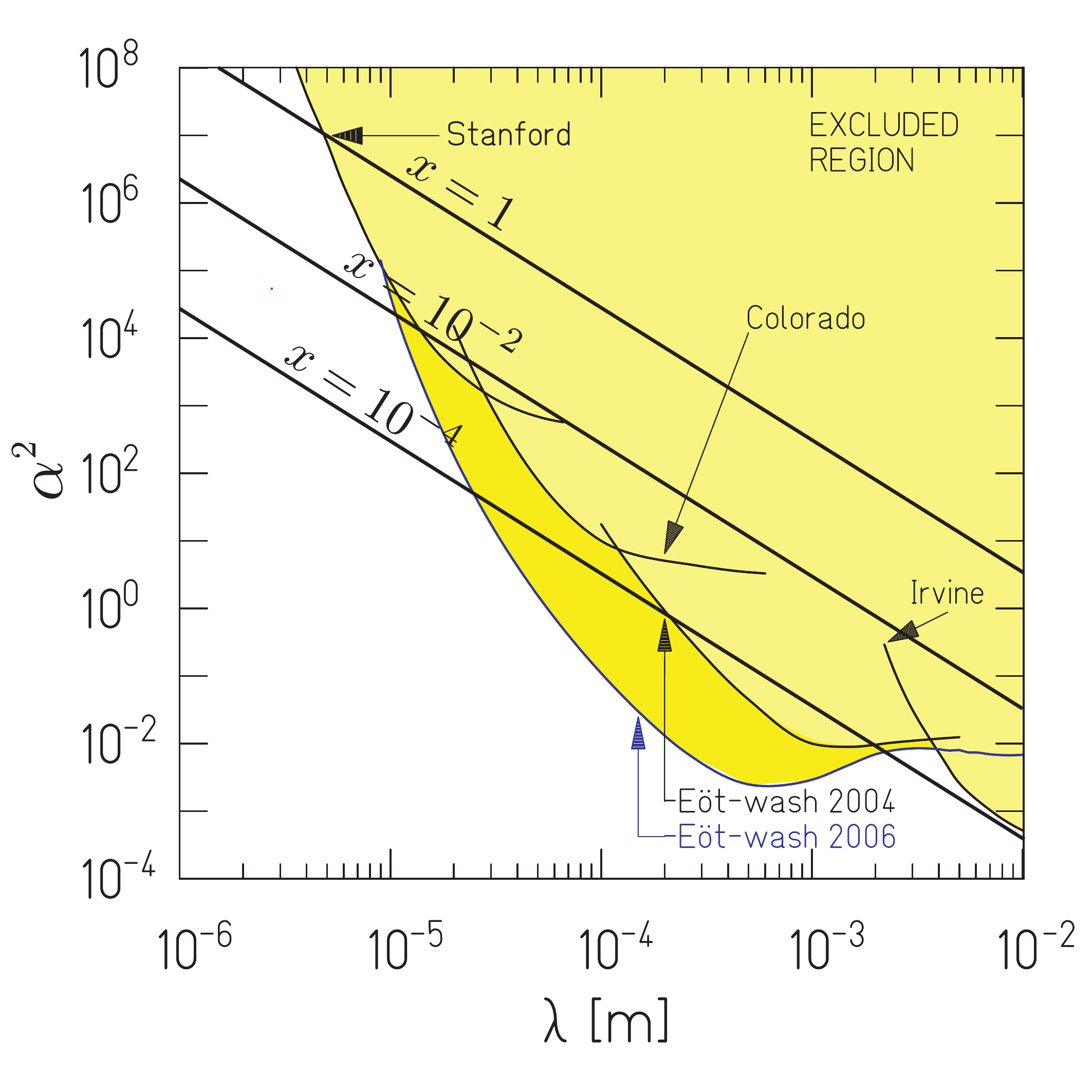}} 
      \end{center} \vspace{-5mm} \caption{We plot the constraints on the mass $m_\varphi$ and coupling $A = x m_\varphi$ coming from fifth force experiments, and taking $g_{h NN}$ to the maximum 
of its allowed range. The range of the force is just $\lambda = m_\varphi^{-1}$. The coupling $\alpha$ is obtained in eq. (\ref{alpha}) by assuming $m_h \simeq 120$ GeV. For two different mass ranges, the lines corresponding to $x=1$, $x=10^{-2}$ and $x=10^{-4}$ are superimposed on the plots of references~\cite{adelberger} (upper panel) and~\cite{kapner} (lower panel).\label{fig3}} 
\end{figure}

In the limit of a very long range force, the value of $\alpha$ is bounded by post-Newtonian tests of General Relativity to  $\alpha^2 \lesssim 10^{-5}$ \cite{bertotti}. However, one can easily see that for mass
range of $m_\varphi$ below $10^{-12}$ eV, the relative strength of the $\phi$-induced force drops 
below $10^{-14}$ from the gravitational field strength, which would make it extremely challenging for 
experimental detection and immune to the Solar System tests. 
Thus, it is more interesting to consider intermediate-range forces. Tests of gravitational inverse-square law limit the Yukawa component of the gravitational potential  \cite{adelberger,kapner}. By means of equation (\ref{alpha}), such tests give a bound on $A$. This is shown in Fig. \ref{fig3}. The two panels are elaborations of plots taken from Refs.~\cite{adelberger} and ~\cite{kapner}. A force with similar values of $m_\varphi$ and $A$ 
($x  \simeq 1$) is excluded in the range of masses $m_\varphi \simeq 10^{-8} {\rm eV} - 10^{-3}$ eV.

The calculations of the EP-violating part of the scalar exchange is a far more delicate excercise. 
One should recognize that the equivalence principle is violated already at the level of nucleons,
that is $g_{h nn}/m_n \neq g_{ h pp}/m_p$. As is well-known, the neutron and proton mass difference comes 
about because of the unequal quark masses, and electromagnetic contribution to the nucleon mass. 
One can estimate $(m_n - m_p) |_{m_u\neq m_d} \simeq 2.1$ MeV and $(m_n - m_p) |_{\rm EM } \simeq -0.8$ MeV,
so that together both contributions combine to the observable mass difference $\Delta m_{np} =1.3$ MeV.
The $\varphi$-dependence of both pieces is completely different. Because of the loop smallness of $g_{h \gamma\gamma}$
the electromagnetic fraction of nucleon mass is far less dependent on $\varphi$:
$ \partial(m_n - m_p) |_{\rm EM }/\partial h   \ll \partial (m_n - m_p) |_{m_u\neq m_d}/\partial h $. Therefore, when we estimate the mass of an atom, we add to the universal term proportional to the baryon number a correction proportional the the nucleon mass difference:
\begin{equation} \label{Mass}
m = (N+Z) m_{\rm nuc}(\varphi) + \frac{N-Z}{2}\Delta_{np}m(\varphi) +\dots
\end{equation}
The first term in (\ref{Mass}) produces the universal coupling $\alpha$ calculated in (\ref{alpha}). The composition-dependent correction reads 
\begin{eqnarray}
\alpha^{\rm EPV} \simeq \alpha \frac{N-Z}{2(N+Z)}\frac{\Delta m_{np}}{m_N}\left(\frac{m_N}{g_{hNN}}
\frac{\partial \Delta m_{np}/\partial h}{\Delta m_{np}} - 1 \right )\nonumber\\
\simeq \alpha \frac{N-Z}{2(N+Z)}\times 3\times 10^{-3} .
\label{EPestimate}
\end{eqnarray}
This may lead to a sizable variation of acceleration $\Delta a$ between light atoms with $Z=N$ and heavy atoms with $\frac{N-Z}{2(N+Z)}\simeq 0.1 $,
\begin{equation}
\frac{\Delta a}{a} \simeq \alpha^2 \times {\cal O}(10^{-3}-10^{-4}) \, .
\end{equation}
Other important effect should be related to the 
dependence of the nuclear binding energy on $\varphi$, 
that can easily reach a level comparable to (\ref{EPestimate}).
More detailed considerations of nuclear mass dependence on 
$\varphi$ go outside the scope of the present paper. 

As long as we adhere to our naturalness condition $A \simeq m_\varphi$, the present bounds on composition dependent EP violations ($\Delta a/a \lesssim 10^{-13}$) are easily evaded. When the Earth is the common attractor of the two free-falling bodies, the relevant range $m_\phi^{-1} \simeq 10^4$ km turns into extremely tiny values for the coupling $A$. Still, if we were to consider more fine-tuned scenarios ($m_\varphi \ll A$), it is interesting to note that a fifth force attached to the Higgs portal displays a peculiar relation between composition independent and composition dependent effects, as clearly follows from eq. (\ref{EPestimate}). In principle, this allows to distinguish between the Higgs portal and, \emph{e.g.}, the string-inspired scenarios \cite{dampol,dpv1,dpv2}.

\section{Cosmological constraints}

Since all couplings to SM particles are suppressed by a factor $Av/m_h^2$, 
the scalar field is sufficiently stable to be a non-thermal relic (for earlier studies of scalar dark matter see, \emph{e.g.},~\cite{Lee,matos}). Its decay rate into photons
is smaller than the current Hubble rate as long as the mass is under a keV:
\begin{equation}
\Gamma_\varphi = \frac{m_\varphi^3A^2 g_{h\gamma\gamma}^2}{4\pi m_h^4} \simeq 
10^{-37}{~\rm eV}\times  x^2 \left(\frac{m_\varphi}{1 {\rm keV}}\right)^5 
\left( \frac{100 ~{\rm GeV}}{m_h}\right)^4.
\end{equation} 
However, as emphasized in \cite{Pospelov}, the constraints from the gamma 
ray background would provide much tighter constraints, and in what follows we will 
concentrate on the sub-eV range. 

The abundance of $\varphi$-particles today can be estimated~\cite{kolb,muk} in terms of the initial misalignment $\varphi_*$ of the field from its minimum at the time $t_*$ when $m_\varphi \sim 3 H$. 
At that moment, the field starts oscillating around the minimum of its potential and behaves like non-relativistic matter. The number of particles in a comoving volume is conserved, so that 
$n_\varphi/s =$ const, where $s = 0.44 g_* T^3$ is the entropy density, $n_\varphi$ is the number density of $\varphi$ particles and $g_*$ is the number of effective degrees of freedom in equilibrium with the photons.  
The (average) energy density of $\varphi$ today is thus given by $\rho^0_\varphi = m_\varphi n_\varphi s^0/s$, where $n_\varphi$ and $s$ should be taken at $t_*$. Unsing the relations $m_\varphi n_\varphi = m_\varphi^2 \varphi_*^2 /2$ and $s^0/s_* = (2/g_*)(T_\gamma^0/T_*)^3$ together with the Hubble rate
\begin{equation} \label{H}
H \simeq {\tilde g}_*^{1/2} \frac{T^2}{3 M_P},
\end{equation}
we express $T_*$ in terms of the parameters of the model to obtain 
\begin{equation} \label{nonthermal}
\Omega_\varphi h^2 = 0.4 \frac{{\tilde g}_*^{3/4}}{g_*} \left(\frac{m_\varphi}{10^{-9} {\rm eV}}\right)^{1/2} \left(\frac{\varphi_*}{10^{14} {\rm GeV}}\right)^2,
\end{equation}
where ${\tilde g}_*$ is the number of degrees of freedom relevant for  estimating $H$. Since 
$T_* \sim 10^5 (m_\varphi/{\rm eV})^{1/2} {\rm GeV}$, in the mass range of interest ${\tilde g}_* = g_*$ and $0.4 \, {\tilde g}_*^{3/4}/{g_*} \simeq O(0.1)$.

An important constraint on the model comes from the smallest allowed mass for a dark matter particle. 
The observations of smallest halos show their size to be comparable to 1 kpc \cite{Willman}, 
which means that the Compton wavelength of $\varphi$ field would have to be comparable or smaller 
that this scale. This in turn imposes the constraint on $\varphi_*$:
\begin{eqnarray}
m_\varphi > 10^{-26}~{\rm eV}~\Rightarrow~ \varphi_* < 2\times 10^{18}~ {\rm GeV} \times 
\left( \frac{\Omega_\varphi h^2 }{0.1} \right)^{1/2}.
\label{massbound}
\end{eqnarray}
Notice that at the boundary of the allowed value, $m_\varphi\sim 10^{-26}$ eV, the oscillations 
start around 10 eV, that is just before the matter-radiation equality.

By eq. (\ref{nonthermal}) it is clear that the abundance of $\varphi$-particles depends on the VEV of the field at the moment when the Hubble parameter becomes of the order of its mass. This is ultimately a matter of initial conditions. However, it is interesting to study the preceding evolution of $\varphi$ up to electroweak symmetry breaking. During inflation, while the field would classically stay constant, its vacuum expectation value 
gets random kicks of order $H/2\pi$ every Hubble time due to quantum fluctuations (see e.g.~\cite{linde}). Its behavior can be described formally with a Langevin type equation~\cite{dpv1}
\begin{equation}
\frac{d \phi}{d p} = \frac{H(p)}{2 \pi} \zeta(p).
\end{equation}
In the above $p = \ln a$ is the number of e-folds and $\zeta$ is a Gaussian random variable. Its $p$-averages are $\langle \zeta(p) \rangle = 0$, $\langle \zeta(p) \zeta(p') \rangle = \delta(p-p')$. It is straightforward to estimate the expected shift in the field during inflation:
\begin{equation}
|\Delta \phi_{\rm inf}| \equiv \sqrt{\langle (\phi_{\rm end} - \phi_{\rm in})^2 \rangle} = \frac{1}{2 \pi}\left(\int_{p_{\rm in}}^{p_{\rm end}} \!\! dp' H^2(p')\right)^{1/2} \!\!.
\end{equation}
As a rough order of magnitude, this gives $\Delta \phi_{\rm inf} \gtrsim 10  H_{\rm CMB}$, where $H_{\rm CMB}$ is the Hubble parameter at the epoch where the scales relevant for the CMB left the horizon. However, in scenarios with a long epoch of self-regenerating inflation,  $\Delta \phi_{\rm inf}$ can be much larger. 

At the onset of radiation domination quantum fluctuations become irrelevant and the field is governed by the classical equation
\begin{equation} \label{bai}
{\ddot \phi} + 3 H {\dot \phi} + m^2_\varphi \phi +  A \langle H^\dagger H \rangle = 0.
\end{equation}
The behavior of $\phi$ up to electro-weak phase transition is obtained by neglecting the second derivative and the mass term from the above equation. While the Higgs field is in thermal equilibrium we have, with good approximation~\cite{muk}, $\langle H^\dagger H \rangle = 3 T^2$.
By using (\ref{H}) we thus get
\begin{equation}
{\dot \phi} {\tilde g}_*^{1/2} = - 3 A M_P\, .
\end{equation}
To a good approximation, the field has a constant velocity, which justifies neglecting the second derivative term in (\ref{bai}). Every time a relativistic species leaves the thermal bath, ${\tilde g}_*$ decreases, giving a little ``kick" to the field's velocity. Thus, the details of this mechanism depend on the physics beyond the Standard Model. By making the minimal assumption ${\tilde g}_* \simeq 100$ we count only for the SM degrees of freedom and obtain a shift in field space. 
\begin{equation} \label{radiation}
\frac{\phi_{\rm EW} - \phi_{\rm end}}{M_P} = - 0.4\ A\, t_{\rm EW} = - 2\, x \times 10^{-6}\frac{m_\varphi}{10^{-10}\rm eV},
\end{equation}
where the subscript EW indicates quantities at electro-weak phase transition.

Finally, at the onset of EW phase transition, the field finds itself displaced from its true minimum by an amount $\varphi_{\rm EW}  \equiv \phi_{\rm EW} - \phi_0$, where $\phi_0$ is given in eq. (\ref{shifts}),
\begin{equation} \label{shift2}
\phi_0 \simeq - 3\, x \times 10^{23}\left(\frac{m_\varphi}{\rm 10^{-10} eV}\right)^{-1}{\rm GeV}.
\end{equation}
A potential disparity between (\ref{massbound}) and (\ref{shift2})
signifies possible fine-tuning problem. The starting point for the $\varphi$ field at the end of inflation
would have to be reasonably close to $\phi_0$, 
Therefore, if we start, say, with $\phi_{\rm end} = 0$ at the end of inflation, the field starts running towards its true minimum, $\phi_0$ thanks to the coupling to the Higgs (\ref{radiation}). However, for masses $m_\varphi \lesssim 10^{-5}$ eV, the shift during radiation domination (\ref{radiation}) is irrelevant with respect to the scale 
set by (\ref{shift2}). If $\varphi$ mediates long/intermediate range ($\lambda \gtrsim$ cm) forces, its initial value $\varphi_*$ has to be fine tuned ($\varphi_* \ll |\phi_0|$), or otherwise $\varphi$-particles are overproduced.

Any super-cold dark matter, such as axion or $\varphi$ field discussed in this 
paper, are prone to the CMB constraints on the amount of isocurvature perturbations
(For the recent discussions of the axion isocurvature perturbations see {\em e.g.} \cite{isoax1,isoax2}). 
Before going to implications of these constraints for the model, we would like 
to comment that the scalar field with the quadratic potential is less susceptible to the 
isocurvature constraints than axions. In case of the quadratic potential 
the increase in the homogeneous displacement from the minimum, $\varphi_*$,
 over the fluctuating value $\delta \phi$ leads to the  $\delta\varphi/\varphi_*$ 
 suppression of the isocurvature perturbations, that in principle can be made 
 arbitrarily small by the increase of $\varphi_*$. In contrast, the 
 increase in the homogenous value of the axion field due to the periodicity of the 
 potential $V_a(a)=V(a+2\pi f_a)$ 
can lead to at most $\delta a/f_a$ suppression of the isocurvature 
 perturbations. 

During inflation, the field $\varphi$ undergoes fluctuations of order
$\delta \varphi = H/2\pi$ as any other light field, $H$ being the Hubble rate at the time when the fluctuation exits the horizon. The produced perturbation are of isocurvature type. Following the standard treatment that also applies to axions~\cite{kolb,komatsu}, we can estimate the power spectrum of entropy perturbations ${\cal P}_{\cal S}(k)$  and compare it to that of curvature perturbations ${\cal P}_{\cal R}(k)$. The ratio of the two defines a parameter
\begin{equation}
\frac{\alpha(k)}{1 - \alpha(k)} \equiv \frac{{\cal P}_{\cal S}(k)}{{\cal P}_{\cal R}(k)} = 8 \epsilon c_s \frac{\Omega_\varphi^2}{\Omega_c^2} \frac{M_P^2}{\varphi_*^2},
\end{equation}
where $\Omega_c$ is proportional to the total energy density in dark matter, $\epsilon$ is the usual inflationary slow-roll parameter and $c_s$ the speed of sound of the adiabatic fluctuations during inflation~\cite{justin}.  

By using (\ref{nonthermal}) and $\Omega_c h^2 \simeq 0.1$ we get, for small $\alpha(k_0)$,
\begin{equation}
\alpha(k) = 4.7 \times 10^9 \epsilon c_s \frac{\Omega_\varphi}{\Omega_c}
\left(\frac{m_\phi} {10^{-9}~\rm eV}\right )^{1/2}
\end{equation}
The above result can nicely be re-expressed in terms of the tensor-to-scalar ratio $r = 16 \epsilon c_s$. At the pivot wavenumber $k_0 = 0.002 \ {\rm Mpc}^{-1}$ the limit set by WMAP+BAO+SN is $\alpha(k_0)<0.067$~\cite{komatsu}. 
This gives the rather strict constraint
\begin{equation} \label{upper}
r \frac{\Omega_\varphi}{\Omega_c} \left(\frac{m_\phi} {10^{-9}~\rm eV}\right )^{1/2}\lesssim 2.3 \times 10^{-10},
\end{equation}
which is very similar to the conclusions reached for the axion cosmology \cite{isoax1,isoax2}. 
If we insist on $\Omega_\varphi$ making most of the cold dark matter density, 
this result shows that the detectable level of inflationary gravitational 
waves ($r > 10^{-2}$) implies a very light scalar, close to the bound (\ref{massbound}),
which would make $\varphi$-mediated fifth force totally negligible. 
Conversely, a detectable level of the fifth force ($m_\varphi >A> 10^{-9}$ eV) would imply 
tiny $r$ on the order of $10^{-10}$ favoring some 
intermediate scale inflationary scenarios, $H \sim O(r^{1/2}\times 10^{14})$ GeV.
Given that in some models of inflation (see {\em e.g.}~\cite{rouzbeh}) the Hubble parameter can be as low as 
$H\sim {\rm GeV}$, producing a tensor to scalar ratio 
$r \simeq 10^{-28}$, constructing an inflationary model
with the Hubble parameter at some intermediate scale does not pause any model-building challenge.

\section{Discussion}

The model we considered in this work is very similar to the linearized version of the 
Brans-Dicke (BD) theory when the scalar field is supplied with the 
mass term. Indeed, the transformation from the Jordan to the Einstein frame 
puts the BD scalar in front of any dimensionful parameter. Therefore, 
the $A$ parameter from the model considered here can be identified with 
$A \sim m_h^2 / (\omega^{1/2} M_{P})$, where $\omega$ is the BD parameter.  
It is very important to keep in mind, however, one crucial difference. 
In the BD theory, the $\phi$-field also couples to all massive states that may 
exist beyond the SM states, and therefore, even at the electroweak scale one should 
expect the extension of Eq. (\ref{potential}) by additional higher-dimensional operators. 
Such terms  alter the couplings of BD scalar to matter, 
and make couplings to gauge bosons, {\em e.g.} $g_{\phi\gamma\gamma}$, different
from the values in the model considered here. 
Moreover, the BD theory requires explicit UV completion, while the model 
with coupling via the super-renormalizable portal assumes 
that higher-dimensional operators are absent from the beginning and generated 
only via the SM loops with the $\varphi$-independent UV cutoff. 

The key feature of the model considered here is its technical naturalness. It allows to 
have a relatively light scalar dark matter that generate medium-range attractive force 
without extra fine tuning of the parameters in the Lagrangian. 
A detectable level of the fifth force, would have to be 
combined with inflationary scenarios with low $r$ and face with the
potential fine-tuning problem in initial condition of the scalar field value. 
One of the most interesting (albeit fine-tuned) scenarios that can have particle physics implications not 
considered in this paper is the $\phi$-dependence of the electroweak phase transition. 
If $m_\varphi$ is taken comparable to the Hubble rate at $T=100$ GeV, $A>m_\varphi$ 
can lead to $|A\phi_*| \sim 10^4$ GeV$^2$, thus altering the properties of the electorweak sector 
close to the phase tansition point. This way, one could change the order of the phase transition, 
and make it first order if the effective Higgs mass is pushed below 50 GeV. 

The model considered here falls into the class of the "super-cool" dark matter models, 
such as axion dark matter. Another example, worth of investigation is the vector 
dark matter. There, the coupling of vector fields to the SM and the mass of the 
vector fields do not have to follow the strength$\times$range=const constraint of the scalar case. 
This could open more room for the fifth-force mediated by the vector-like 
sub-eV dark matter. 

We would like to thank N. Afshordi, N. Barnaby, J. Bond, C. Burgess, A. Erickcek and A. Nicolis for useful discussions. Research at the Perimeter
Institute supported in part by the Government of Canada
through NSERC and by the Province of Ontario through MEDT.

\end{document}